# Usefulness of altmetrics for measuring the broader impact of research: A case study using data from PLOS (altmetrics) and F1000Prime (paper tags)

Lutz Bornmann

Division for Science and Innovation Studies

Administrative Headquarters of the Max Planck Society

Hofgartenstr. 8,

80539 Munich, Germany.

Email: bornmann@gv.mpg.de

**Abstract**

Purpose: Whereas citation counts allow the measurement of the impact of research on research itself, an important role in the measurement of the impact of research on other parts of society is ascribed to altmetrics. The present case study investigates the usefulness of altmetrics for measuring the broader impact of research.

Methods: This case study is essentially based on a dataset with papers obtained from F1000. The dataset was augmented with altmetrics (such as Twitter counts) which were provided by PLOS (the Public Library of Science). In total, the case study covers a total of 1,082 papers.

Findings: The F1000 dataset contains tags on papers which were assigned intellectually by experts and which can characterise a paper. The most interesting tag for altmetric research is "good for teaching". This tag is assigned to papers which could be of interest to a wider circle of readers than the peers in a specialist area. Particularly on Facebook and Twitter, one could expect papers with this tag to be mentioned more often than those without this tag. With respect to the "good for teaching" tag, the results from regression models were able to confirm these expectations: Papers with this tag show significantly higher Facebook and Twitter counts than papers without this tag. This association could not be seen with Mendeley or Figshare counts (that is with counts from platforms which are chiefly of interest in a scientific context).

Conclusions: The results of the current study indicate that Facebook and Twitter, but not Figshare or Mendeley, can provide indications of papers which are of interest to a broader circle of readers (and not only for the peers in a specialist area), and seem therefore be useful for societal impact measurement.

# 1 Introduction

Since the research department of governments competes for the allocation of public funding with other departments (such as defence), governments need to be informed about the benefits (such as the impact) of research (in order to be able to balance funding between departments). Whereas scientometricians have chiefly used citation scores as an instrument for the measurement of the impact of research, in recent years altmetrics (short for alternative metrics) has provided a further attractive possibility for measuring the impact of research. "Altmetrics refers to data sources, tools, and metrics (other than citations) that provide potentially relevant information on the impact of scientific outputs (e.g., the number of times a publication has been tweeted, shared on Facebook, or read in Mendeley). Altmetrics opens the door to a broader interpretation of the concept of impact and to more diverse forms of impact analysis" (Waltman and Costas, 2014). Whereas citations allow a measurement of the impact of research on research itself, the assumption is that altmetrics can also capture the impact beyond the realm of science.

There are already a number of studies concerning altmetrics in scientometrics. An overview of these studies can be found in, for example, Bar-Ilan et al. (2014), Bornmann (2014a), Haustein (2014), and Priem (2014). An overview of the various available altmetrics can be found in Priem and Hemminger (2010). For example, Twitter (www.twitter.com) is the best-known microblogging application, whose data can be used as an altmetric. This application allows the user to post short messages (tweets) of up to 140 characters. "These tweets can be categorised, shared, sent directly to other users and linked to websites or scientific papers" (Darling et al., 2013). Other sources of altmetrics are Mendeley, Facebook, and Figshare. Although several empirical studies have been conducted on altmetrics, it is still not clear what can actually be measured with altmetrics: Is it really the hoped-for broad impact on society? This question is investigated in the current case study – following the study design of Bornmann (2014c).

In January 2002, a new type of post-publication peer-review system has been launched, in which around 5000 Faculty members are asked "to identify, evaluate and comment on the most interesting papers they read for themselves each month – regardless of the journal in which they appear" (Wets et al., 2003). The faculty members also attach tags to the papers indicating their relevance for science (e.g. "new finding"), but which can also serve other purposes. A tag which is of particular interest for altmetrics research, is "good for teaching". Papers can be marked by Faculty members in this way if they represent a key paper in a field, are well written, provide a good overview of a topic, and/or are well suited as literature for students. Thus, one can expect that tagged papers are of special interest for and can be understood by people who are not researchers (within a specific field). In other words, papers marked with this tag can be expected to have an impact beyond science itself, unlike papers without this tag. This expectation will be investigated in this case study with various altmetrics data.

The present study is essentially based on a dataset with papers (and their assessments and tags from faculty members) extracted from F1000. This dataset is supplemented with altmetric data (e.g. Twitter counts) for papers published by PLOS (the Public Library of Science, a non-profit publisher). The downloadable CSV file is available via Figshare (10.6084/m9.figshare.1189396). PLOS regularly publishes a large number of article level metrics (ALMs) relating to the journals published by PLOS. ALMs at PLOS do not only contain altmetrics, but also other metrics, such as citations from Scopus and downloads. Using the F1000 tags (especially the "good for teaching" tag) and PLOS ALMs, this case study explores the question whether altmetrics are able to measure some kind of societal impact.

The results of this case study will be compared with those of Bornmann (2014c) who used the same F1000 dataset matched with an altmetrics dataset from Altmetric (http://www.altmetric.com/) – a start-up that focuses on making ALMs available. Whereas the PLOS dataset offers a wide range of ALMs, Bornmann (2014c) could only include two altmetrics in this case study: the total altmetric score and Twitter counts.

## 2   Methods

**2.1   Peer ratings provided by F1000**

F1000 is a peer review system (ex-post) of papers from biological and medical journals. F1000 is part of the Science Navigation Group. This group of independent companies publishes and develops information services for the biomedical community. F1000 started with F1000 Biology, which was launched in 2002. In 2006, F1000 Medicine was added. Both services were merged in 2009 and constitute the F1000 database now. Papers for F1000 are selected by a global "Faculty" of leading scientists and clinicians who were nominated by peers. The Faculty nowadays numbers more than 5,000 experts worldwide, assisted by further associates, which are organised into more than 40 subjects. Faculty members rate the papers and explain their importance (F1000, 2012). Faculty members can choose and evaluate any paper that interests them. Since only a restricted set of papers from the medical and biological journals covered is reviewed, most of the papers are actually not (Wouters and Costas, 2012, Kreiman and Maunsell, 2011). Although many papers published in popular and high-profile journals (e.g. *Nature*, *New England Journal of Medicine*, *Science*) are evaluated, 85% of the papers selected come from specialised or less well-known journals (Wouters and Costas, 2012).

The papers selected for F1000 are rated by the Faculty members as "Good," "Very good" or "Exceptional" which is equivalent to scores of 1, 2, or 3, respectively. In many cases a paper is assessed not just by one member but by several. According to the results of Bornmann (in press) "the papers received between one and 20 recommendations from different Faculty members. Most of the papers (around 94%) have one recommendation (around 81%) or two recommendations (around 13%) by Faculty members." The FFa (F1000 Article Factor), given as a total score in the F1000 database, is a sum score from the different recommendations for a publication. Besides making recommendations, faculty members also tag publications with classifications, as for example (see http://f1000.com/prime/about/whatis/how):

- Clinical Trial (non-RCT): investigates the effects of an intervention (but neither randomised nor controlled) in human subjects.
- Confirmation: the findings of the article validate previously published data or hypotheses.
- Controversial: findings of the article either challenge the established dogma in a given field, or require further validation before they can be considered irrefutable.
- Good for Teaching: a key article in that field and/or a particularly well written article that provides a good overview of a topic or is an excellent example of which students should be aware. The "good for teaching" tag is relatively new for F1000Prime; it was introduced only in 2011.
- Interesting Hypothesis: proposes a novel model or hypothesis that the recommending Faculty Member found worthy of comment.
- New Finding: presents original data, models or hypotheses.
- Novel Drug Target: the article suggests a specific, new therapeutic target for drug discovery (rather than a new drug).
- Refutation: the findings of the article disprove previously published data or hypotheses (where a specific finding is being refuted, a reference citation is normally required).

- Technical Advance: introduces a new practical/theoretical technique, or novel use or combination of an existing technique or techniques.

The classifications of the papers are intended to be an additional filter rather than part of the rating by a Faculty member. The tags are very useful because they are assigned by an expert. Whereas literature databases (e.g. Web of Science, WoS, Thomson Reuters) cannot be searched for negative results, or clinical practice changing papers, the human expert-assigned tags enable this in F1000Prime.

Overall, the F1000 database is regarded as an aid for scientists to receive pointers to the most relevant papers in biomedicine, but also as an important tool for research evaluation purposes. According to Wouters and Costas (2012) "the data and indicators provided by F1000 are without doubt rich and valuable, and the tool has a strong potential for research evaluation, being in fact a good complement to alternative metrics for research assessments at different levels (papers, individuals, journals, etc.)" (p. 14). Thus, it is no wonder that several empirical studies have been published which are based on the F1000 database (Du et al., Li and Thelwall, 2012, Mohammadi and Thelwall, 2013, Waltman and Costas, 2014, Wardle, 2010).

## 2.2 Construction of the dataset

In January 2014, F1000 provided the author with data on all recommendations (and classifications) made and the bibliographic information for the corresponding papers in their system (n=149,227 records). By using the DOI, this dataset was matched with another dataset downloaded from the PLOS homepage (see http://article-level-metrics.plos.org/plos-alm-data/). The PLOS dataset (alm_report_2014-03-10.csv) includes for all papers (n=114,093 records) published by PLOS, a large number of article-level metrics (ALMs) which range from traditional citations (e.g. from Scopus, Elsevier) up to Twitter counts. An overview of the numerous ALMs provided by PLOS can be found in Fenner (2013). Matching of the two datasets with the DOI led to a matched dataset of 3,928 records. The marked reduction in the records with the matching of the datasets can be explained by the high selectivity of F1000 faculty members in the selection of papers, and that the PLOS database contains only a small subset of the journals which the members generally access. The 3,928 records relate to the publication years 2003 to 2013. Since the "Good for teaching" tag, which Faculty members can assign and which has a great importance in the evaluation of altmetrics (see above), was not used by F1000 before 2011, the case study can only include records relating to the years 2012 and 2013. This reduces the dataset once more, from 3,928 records to 1,204 records. Since papers which received more than one recommendation from a faculty member appear multiply in the dataset, the 1,204 records refer to 1,082 papers.

**Table** 1
Proportion of PLOS Papers, which were published after 2011 and which have non-zero counts for the ALMs used (n=1,082)

| ALM | Proportion of papers in 2012 with non-zero counts (n=586) | Proportion of papers in 2013 with non-zero counts (n=496) |
|---|---:|---:|
| **Social Network** | | |
| Facebook | 49.5 | 51.0 |
| Twitter | 49.3 | 62.3 |
| **Usage** | | |
| CiteULike | 31.1 | 20.2 |
| Figshare | 83.6 | 85.3 |
| Mendeley | 84.8 | 47.4 |
| **Citation** | | |
| CrossRef | 94.7 | 64.9 |
| DataCite | 0.3 | 0.4 |
| PubMed Central | 100.0 | 100.0 |
| Scopus | 76.1 | 56.9 |
| **Comments, Blogs, and Media** | | |
| Nature.com posts | 1.4 | 0.0 |
| PLOS comments | 17.2 | 13.9 |
| Reddit | 0.5 | 0.0 |
| Research Blogging | 2.0 | 1.0 |
| ScienceSeeker | 0.0 | 0.0 |
| Wikipedia | 5.5 | 3.4 |
| WordPress | 6.8 | 6.9 |
| **Downloads** | | |

| PLOS downloads | 100.0 | 100.0 |
| PubMed downloads | 81.2 | 44.6 |
| **Recommendations** | | |
| F1000 summary metric | 100.0 | 100.0 |

In Table 1 the most important ALMs from PLOS are listed for all papers included in the study. The dataset with the PLOS ALMs is published at regular intervals, while the listed ALMs are changed. Table 1 therefore relates to the core set of metrics which have generally been listed by PLOS for a longer time (with some exceptions, like DataCite). For each ALM, the table gives the proportion of papers in 2012 and 2013 which have non-zero counts. As the results in the table show, the share varies greatly with the ALMs and ranges from 0% with ScienceSeeker right up to 100% with the PLOS downloads. Under the label "social network" and "usage", Table 1 provides the group of those altmetrics which belong to well-known altmetrics and are in the focus of this case study. Unlike another group of altmetrics listed in the table under "comments, blogs and media", the social network and usage altmetrics are not affected by an inflation of zeros which lead to problems with the statistical analysis (calculation of regression models) (Long and Freese, 2006). Count response models with far more zeros than expected result in incorrect parameter estimates as well as biased standard errors (Hardin and Hilbe, 2012). Of the five social network and usage metrics named in Table 1, four are included in the current study: Facebook, Figshare, Mendeley, and Twitter. CiteULike is disregarded in the case study, since it is a platform similar to Mendeley, and Mendeley generally covers the literature more thoroughly than CiteULike (Bornmann, 2014a).

The four altmetrics included in the case study are outlined in the following. More comprehensive descriptions can be found in the overview works on altmetrics referred to in the introduction. Facebook (facebook.com) is the most often used social media site, where users can create a profile page and interact with other users (Bik and Goldstein, 2013). Since these exchanges with other users may include references to publications, there is the possibility of counting these references and using them as an altmetric. Figshare (figshare.com) is a relatively new application (since 2011), in which users can store their research output (such as papers, data, tables and graphics) in a repository and exchange among each other. Since Figshare tracks the download statistics for the research output listed, these data are used as a source for altmetrics (Hahnel, 2013).

Online reference managers combine social bookmarking service and reference management functionalities in one platform. They can be seen as the scientific variant of social bookmarking platforms, in which users can save and tag web resources (e.g. blogs or web sites). The best-known online reference manager is Mendeley (mendeley.com), which was launched in 2008 (Li et al., 2012). The platform allows users to save or organise literature, to share literature with other users, as well as to save keywords and comments on a publication (or to assign tags to them) (Bar-Ilan et al., 2014, Haustein et al., 2014b). As the source for altmetrics, user counts are taken, which provide the number of readers of publications via the saves of publications (Li et al., 2012).

With micro-blogging, users send short messages to other users of a platform. The best-known microblogging platform is Twitter (twitter.com), which was founded in 2006. Twitter allows the sending of short messages (called tweets). The frequency of the tweets referring to a particular paper can be determined, and used as a source for altmetrics.

## 2.3  Statistical procedure and software used

The statistical software package Stata 13.1 (http://www.stata.com/) is used for this case study; in particular, the Stata commands nbreg, margins, and coefplot are used.

For the statistical analyses of the data, a series of regression models has been estimated. Since the outcome variables (Facebook, Figshare, Mendeley, and Twitter counts) in the models indicate "how many times something has happened" (Long and Freese, 2006), they are count variables. The Poisson distribution is as a rule used to model information on counts. However, this distribution rarely fits in the statistical analysis of altmetric data, due to overdispersion. "That is, the [Poisson] model underfits the amount of dispersion in the outcome" (Long and Freese, 2006). The standard model to account for overdispersion is the negative binomial (Hausman et al., 1984). Thus, negative binomial regression models are calculated in the present case study (Hilbe, 2007).

The violation of the assumption of independent observations by including several F1000 recommendation scores associated with a paper is considered in the regression models by using the cluster option in Stata (StataCorp., 2013). This option specifies that the scores are independent across papers but are not necessarily independent within the same paper (Hosmer and Lemeshow, 2000, section 8.3).

The publication years (2012 and 2013) of the papers were included in the models predicting different counts as exposure time (Long and Freese, 2006). The exposure option provided in Stata takes into account the time that a paper is available for mentions.

In this case study, predicted probabilities are used to make the results easy to understand and interpret. Such predictions are referred to as margins, predictive margins, or adjusted predictions (Bornmann and Williams, 2013, Williams, 2012). The predictions allow an interpretation of the empirical results which goes beyond the statistical significance test. Whereas adjusted predictions can provide a practical feel for the practical significance of the findings, the regression models illustrate which effects are statistically significant and what the direction of the effects is.

# 3 Results

## 3.1 The distribution and selection of the tags in the dataset

Table 2 shows the distribution of the tags across the records in the dataset (with multiple occurrences of PLOS papers) or total tag mentions (see "total" line). It is very clear that the tags were applied very differently: Whereas, for example, "new finding" makes up about half of the tag mentions, for "controversial" it is only about 5%. In order to be able to make a <u>reliable</u> statement about the validity of the altmetrics, the following statistical analysis does not include all tags, but only those with more than 5% of mentions or allocated to more than 10% of records (see grey marked tags).

**Table** 2
Tags allocated by faculty members (n=1,204 records, n=1,787 tag mentions).

| Tag | Absolute numbers | Percent of tag mentions | Percent of records |
| --- | --- | --- | --- |
| New finding | 823 | 46.05 | 68.36 |
| Interesting hypothesis | 249 | 13.93 | 20.68 |
| Confirmation | 197 | 11.02 | 16.36 |
| Good for teaching | 163 | 9.12 | 13.54 |
| Technical advance | 159 | 8.90 | 13.21 |
| Controversial | 88 | 4.92 | 7.31 |
| Novel drug target | 69 | 3.86 | 5.73 |
| Refutation | 13 | 0.73 | 1.08 |
| Systematic review | 10 | 0.56 | 0.83 |
| Review | 8 | 0.45 | 0.66 |
| Negative | 4 | 0.22 | 0.33 |
| Clinical trial (non-RCT) | 4 | 0.22 | 0.33 |
| Total | 1787 | 100.00 | 148.42 |

What expectations are there in the current study in relation to the connection between altmetrics counts and the categorisation of papers with the five selected tags (which are described in further detail in section 2.1)? In connection with "new finding", "confirmation" and "interesting hypothesis", it is expected that the Figshare and Mendeley counts for such papers would be higher for those where a faculty member has used this tag than for those where this did not happen. These tags chiefly relate to aspects which are relevant in a scientific context, and Figshare and Mendeley are mainly used in this context. With Facebook and Twitter counts, this difference between tagged or untagged papers is not expected since the two platforms do not appeal to a specifically scientific set of users.

We can expect that papers tagged with "good for teaching" would (also) be interesting for a set of people outside science (or research). These are papers, which are well-written, provide an overview of a topic, and are well suited for teaching. Therefore, higher Facebook and Twitter counts are expected for papers with this tag than for papers without it. With the Mendeley and Figshare counts, this difference is not expected, since these platforms are more strongly oriented to science.

The "technical advance" tag is used on papers that present a new technique or tool (whether that's a lab technique/ tool or a clinical one) that make an advance on an existing technique. The tag can be used both for research papers and outside, i.e. clinical or fieldwork. Accordingly, no great count difference between the four altmetrics is to be expected.

**3.2 How do the altmetric counts differ for differently tagged papers?**

In order to examine how Mendeley, Facebook, Twitter and Figshare counts differ with differently tagged papers, four regression models with four counts as dependent variable and the tags as independent variable were calculated. As Table 3 shows, each model includes the individual recommendation scores of the faculty members alongside the tags. This can be used to investigate the influence of the tags on the different counts – under the control of the effects of the recommendations. Since the recommendations reflect the quality of the papers, the results of the tags are adjusted for the quality of the papers. In other words: the different results for the tags can hardly be traced back to the differing quality of the papers. The model also includes the journals in which the papers were published. This allows the influence of the journal on the counts to be controlled for in the analysis. Thus, for instance, one may expect high-impact journals (e.g. PLOS Biology) to have a larger circle of readers than journals with a lower impact, and that their contributions would also be mentioned more often in social media than the contributions of other journals (Haustein and Siebenlist, 2011).

**Table** 3
Dependent and independent variables which were included in the four negative binomial regression models

| Variable | Mean/ Percent | Standard deviation | Minimum | Maximum |
|---|---|---|---|---|
| **Dependent variables** | | | | |
| Mendeley counts (model 1) | 14.87 | 34.82 | 0 | 460 |
| Facebook counts (model 2) | 31.45 | 341.54 | 0 | 11,459 |
| Twitter counts (model 3) | 7.82 | 32.79 | 0 | 800 |
| Figshare counts (model 4) | 7.38 | 12.69 | 0 | 174 |
| **Dependent variables** | | | | |
| Tag | | | | |
| New finding | 68% | | 0 | 1 |
| Confirmation | 16% | | 0 | 1 |
| Interesting hypothesis | 21% | | 0 | 1 |
| Good for teaching | 14% | | 0 | 1 |
| Technical advance | 13% | | 0 | 1 |
| Recommendation of Faculty member | | | | |
| 1 "good" (reference category) | 51% | | 0 | 1 |
| 2 "very good" | 42% | | 0 | 1 |
| 3 "exceptional" | 7% | | 0 | 1 |
| Journal | | | | |
| PLOS Biology (reference category) | 19% | | 0 | 1 |
| PLOS Computational Biology | 3% | | 0 | 1 |
| PLOS Genetics | 12% | | 0 | 1 |
| PLOS Medicine | 3% | | 0 | 1 |
| PLoS Neglected Tropical Diseases | 2% | | 0 | 1 |
| PLoS One | 60% | | 0 | 1 |
| PLoS Pathogens | 1% | | 0 | 1 |
| Number of recommendations | n=1,204 | | | |
| Number of papers | n=1,204 | | | |

The results of the regression models are shown in Table 4. The predicted numbers of counts resulting from the models are presented in Figure 1 for the different recommendation scores and in Figure 2 for the different tags. Since the predicted numbers of counts depend on the models with all independent variables, they are calculated for the different tags under control of the recommendation scores (and thus adjusted for quality). As the results of the regression models in Table 4 show, the coefficients for the recommendation scores of the faculty members are statistically significant with the models for the Mendeley and Twitter counts: Papers assessed as "very good" show higher counts than those just rated as "good". The relation between the different recommendation scores and the predicted numbers of counts is represented in Figure 2. As the graphic shows, with Mendeley, Facebook and Twitter a difference is indeed visible between "good" and "very good", but not between "very good" and "exceptional". The predicted numbers of Figshare counts has almost no relationship – according to the results in Figure 2 – with the recommendation scores.

**Table** 4
Results of four negative binomial regression models

|  | Model 1: Mendeley | Model 2: Facebook | Model 3: Twitter | Model 4: Figshare |
|---|---|---|---|---|
| **Tag** | | | | |
| New finding | 0.05 | 0.50* | -0.14 | 0.21 |
|  | (0.36) | (1.98) | (-0.68) | (1.72) |
| Confirmation | 0.10 | 0.32 | 0.33 | -0.15 |
|  | (0.72) | (0.94) | (1.16) | (-1.22) |
| Interesting hypothesis | -0.19 | 0.35 | 0.08 | 0.11 |
|  | (-1.48) | (1.57) | (0.45) | (0.87) |
| Good for teaching | 0.07 | 0.67** | 0.80*** | 0.23 |
|  | (0.40) | (2.68) | (3.47) | (1.76) |
| Technical advance | 0.58** | 0.10 | 0.40 | 0.05 |
|  | (3.20) | (0.42) | (1.52) | (0.31) |
| **Recommendation of Faculty member** | | | | |
| 1 "good" (reference category) | | | | |
| 2 "very good" | 0.24* | 0.43 | 0.46* | -0.01 |
|  | (2.23) | (1.79) | (2.39) | (-0.05) |
| 3 "exceptional" | 0.21 | 0.46 | 0.32 | -0.12 |
|  | (1.01) | (1.28) | (0.98) | (-0.63) |
| **Journal** | | | | |
| PLOS Biology (reference category) | | | | |
| PLOS Computational Biology | 0.30 | -0.39 | -0.72* | -0.17 |
|  | (0.45) | (-0.65) | (-2.05) | (-0.64) |
| PLOS Genetics | -0.50 | -1.37*** | -1.07*** | -0.19 |
|  | (-1.90) | (-3.83) | (-4.48) | (-0.98) |
| PLOS Medicine | -0.95* | 0.23 | 0.81** | -0.46 |
|  | (-1.99) | (0.40) | (2.61) | (-1.38) |
| PLoS Neglected Tropical Diseases | -0.98** | 3.57*** | 1.17 | -0.22 |
|  | (-2.79) | (3.35) | (1.23) | (-0.74) |
| PLoS Pathogens | -0.89*** | -0.40 | -0.61 | -0.09 |
| | (-3.43) | (-0.75) | (-1.71) | (-0.43) |
| PLoS One | -1.15*** | -0.36 | -0.82*** | -0.23 |
| | (-4.90) | (-1.03) | (-3.54) | (-1.33) |
| Constant | -4.35*** | -5.02*** | -5.51*** | -5.61*** |
| | (-17.26) | (-12.68) | (-21.47) | (-26.96) |

Notes.
$t$ statistics in parentheses
$^{*} p < 0.05$, $^{**} p < 0.01$, $^{***} p < 0.001$

In the Facebook and Twitter models (models 2 and 3), the coefficient for "good for teaching" is statistically significant (see Table 4). Correspondingly, Figure 2 shows higher predicted numbers of counts for papers where this tag is set than for those papers where this was not the case. For example, we can expect a paper with this tag to have around nine Twitter citations more than one without – if the paper is rated as "very good" by Faculty members and has no other tags. This result is in agreement with the results of Bornmann (2014c). The result of the present study for "good for teaching" correspond to the expectations (see above) and indicate that Facebook and Twitter data can indicate papers which are of interest outside the field of science. Since this effect of the "good for teaching" tag is not demonstrable with Figshare and Mendeley data in this form, such indications are not to be expected with these altmetrics.

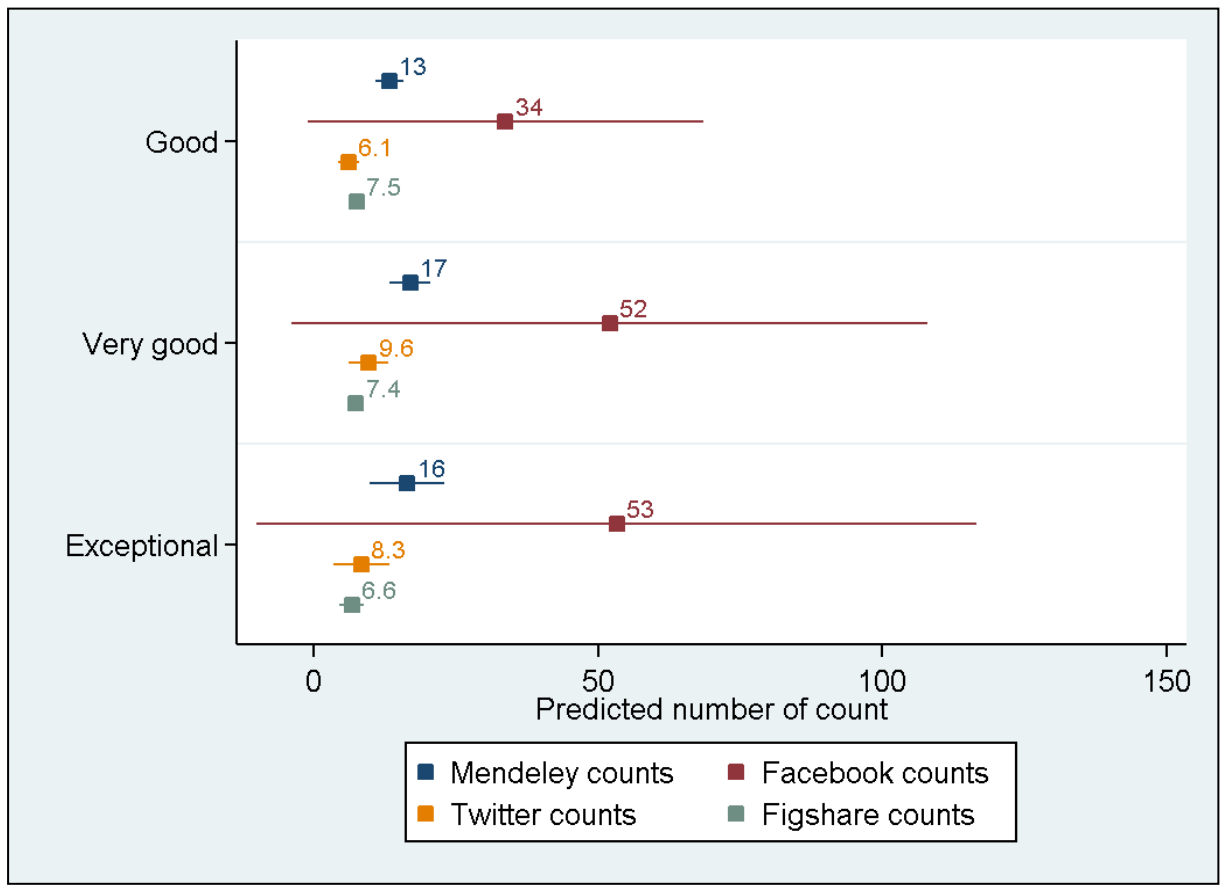

Figure 1. Predicted numbers of Mendeley, Facebook, Twitter, and Figshare counts with 95% confidence intervals for three individual recommendation scores

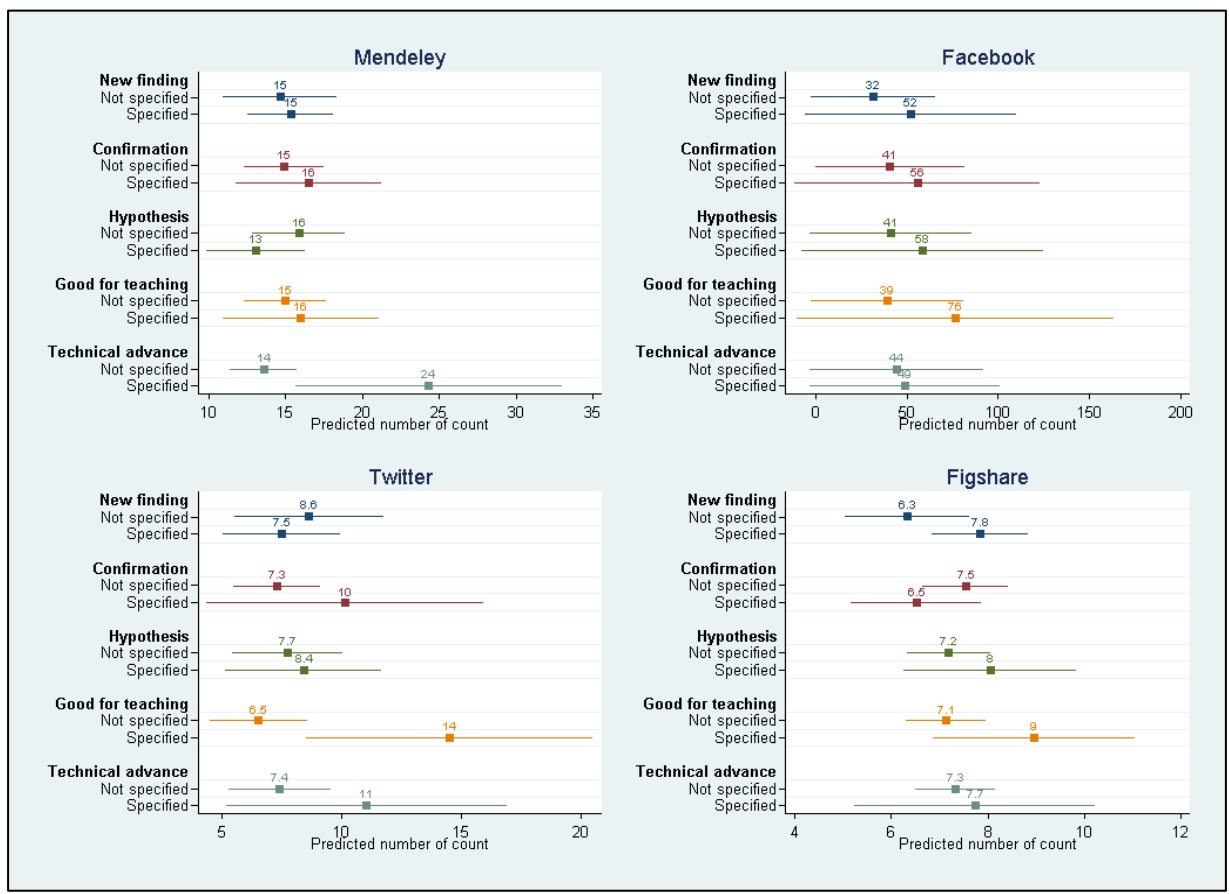

Figure 2. Predicted numbers of Mendeley, Facebook, Twitter, and Figshare counts with 95% confidence intervals for differently tagged papers

Besides the two statistically significant results with the "Good for teaching" tag in the Facebook and Twitter models, the following two statistically significant results appeared in the four regression models (see Table 4): Saves by Mendeley users are particularly to be expected when a paper introduces a new practical/ theoretical technique (tag: "technical advance"). Since these techniques are an important foundation for work in science, it is understandable that papers with this background should be preferentially saved by Mendeley users. "New finding" is statistically significant in the model with which the Facebook counts are evaluated - although Facebook is mainly a platform for people outside of research. New findings therefore do not lead only to higher traditional citation counts, as Bornmann (2014c) shows, but also to higher Facebook mentions. With reference to Figshare counts it is interesting to note that none of the five tags can be associated with markedly higher or lower altmetric counts (see Figure 2). Downloads from the research output stored on Figshare therefore take place relatively independently of the characteristics of the paper (investigated here).

## 4 Discussion

Altmetrics are regarded as an attractive possibility for filling the requirement for indicators for broad impact measurement. The data accessibility of altmetrics is already rather good: Several organisations (such as ImpactStory and Altmetric) have already been founded to collect and provide altmetrics. In addition, altmetrics can also measure the impact of research products other than publications, such datasets, software or artworks. As opposed to traditional citations, many altmetrics (such as Twitter counts) have the advantage that they permit an impact measurement relatively quickly after the publication of a work. Whereas traditional citations allow a measurement of impact at least three years after the appearance of a publication (Bornmann and Marx, 2014), this is already possible in a (significantly) shorter time scale with many altmetrics. For example, Twitter citations on a paper often decline after only a few days.

Since the question which form of impact can be measured with altmetrics has not yet been settled today, this case study addresses the question with a matched dataset of F1000 and PLOS data. The F1000 dataset contains tags on papers which were assigned intellectually by experts and which can characterise a paper. The most interesting tag for altmetric research is "good for teaching". This tag is assigned to papers which could be of interest to a wider circle of readers than the peers in a specialist area. Particularly on Facebook and Twitter, one could expect papers with this tag to be mentioned more often than those without this tag. Facebook and Twitter are particularly in use by people outside the area of science: Although both platforms belong to the most often used social media, it is generally assumed that only a few scientists are actually professionally active there (Darling et al., 2013, Mahrt et al., 2012).

Since the quality of a paper, and not only the contents is expected to play a role in its being mentioned on a platform, the quality of a paper should be controlled for in a corresponding statistical analysis of altmetric counts.

With respect to the "good for teaching" tag, the results of the current study were able to confirm these expectations: Papers with this tag show significantly higher Facebook and Twitter counts than papers without this tag. This association could not be seen with Mendeley or Figshare counts (that is with counts from platforms which are chiefly of interest in a scientific context). Facebook and Twitter thus seem in fact to indicate papers which are of interest for a wider circle of readers. These results are in agreement with those of Bornmann (2014c) who investigated also Twitter counts using the F1000 dataset. In the results from the Mendeley and Figshare platforms it is interesting to note that papers with the "technical advance" tag have significantly more Mendeley counts than papers without this tag. It seems that interesting new techniques or tools for research which are presented in papers or used for a study can be identified with the help of Mendeley counts.

The current study makes a contribution to the clarification of the question of the broad impact of altmetrics. Even if the results with regard to the use of Twitter and Facebook counts are promising, there are a series of unanswered questions which result from the findings presented here: Why do papers with the "good for teaching" tag receive higher Twitter and Facebook counts than papers without this tag? How does the impact vary with time for papers with this tag: When is the peak for the Facebook and Twitter counts reached (i.e. how soon after publication of a paper can one arrive at a broader impact measurement)? The "controversial" tag could not be included in the present study, because the case number was too low. Controversial papers are relevant mostly for the scientific community (as they belong to the scientific debate). It would be interesting to explore whether they have also a strong visibility in Twitter.

There are still many research questions on altmetrics to be answered which go well beyond those mentioned here. An overview of the numerous further questions may be found in NISO Alternative Assessment Metrics Project (2014), Priem (2014) and Haustein (2014).

Although the present study produced new insights into the broad impact measurement using altmetrics, the study is not without limitations:

(1) The study is based on a very specific subset of papers. The sample is rather limited (only PLOS papers filtered by F1000, n=1082). F1000 is a tool focusing on the biomedical fields and only a very small fraction of all biomedical articles are actually covered. For example, the results show that there are Facebook and Twitter mentions for around 50% (or more) of the papers. These coverage figures are remarkably high for sources that have been reported to have a very low coverage in other – significantly larger – samples (Zahedi et al., 2014, Haustein et al., 2014a, Robinson-García et al., 2014).

(2) Another methodological problem is that the case study is a non-zero analysis. Basically all papers have F1000 scores. However, we don't know if there are papers that could be "good for teaching" and are not picked up by F1000. It is not possible to explore whether these "other" papers would have more Twitter activity than the rest.

(3) The most important information in this study is the "good for teaching" tag. Papers are marked in this way if they represent a key paper in a field, are well written, provide a good overview of a topic, and/or are well suited as literature for students. Papers marked with this tag can be expected to have an impact beyond science itself, unlike papers without this tag. Although papers marked by this tag have indeed higher Twitter and Facebook counts, it is not clear whether this impact is beyond science. Twitter, Facebook and other social media are obviously popular outside of science. However, it could not be investigated with the present dataset, whether these people (e.g. journalists or politicians) have been really reached.

4) The fourth limitation concerns the Twitter data used in this study: as PLOS started to collect them routinely in June 2012, many tweets from January to May 2012 will not have been captured.

# 5 Conclusions

Initiated by the wish of governments for a broader impact measurement of scientific activities, the field of scientometrics is currently undergoing a process of deep-seated change, which is described by Bornmann (2014b) as a scientific revolution. An impact measurement can no longer involve only traditional citations, but also indicators which can measure the impact of research in other sectors of society (such as the economy, culture or health services). The results of the current study indicate that Facebook and Twitter, but not Figshare or Mendeley, can provide indications of papers which are of interest to a broader circle of readers (and not only for the peers in a specialist area), and seem therefore be useful for societal impact measurement.

## Acknowledgements

I would like to thank Adie Chan, Ros Dignon and Iain Hrynaszkiewicz from F1000 for providing me with the F1000Prime data set.